\documentclass[sigconf]{acmart}
\usepackage{caption}
\usepackage{subcaption}

\usepackage{xurl}
\usepackage{xtab}
\usepackage{longtable}
\usepackage{subcaption}
\usepackage{array}
\usepackage{tabularx}
\usepackage{enumitem}

\copyrightyear{2021}
\acmYear{2021}
\setcopyright{acmlicensed}
\acmConference[AIES '21] {Proceedings of the 2021 AAAI/ACM Conference on AI, Ethics, and Society}{May 19--21, 2021}{Virtual Event, USA.}
\acmBooktitle{Proceedings of the 2021 AAAI/ACM Conference on AI, Ethics, and Society (AIES '21), May 19--21, 2021, Virtual Event, USA}
\acmPrice{15.00}
\acmISBN{978-1-4503-8473-5/21/05}
\acmDOI{10.1145/11952.107}
\renewcommand{\shortauthors}{Chi et al.}

\begin{document}
\fancyhead{}

\title{Reconfiguring Diversity and Inclusion for AI Ethics}


\author{Nicole Chi}
\affiliation{%
  \institution{Mobius Project}
    \country{USA}}
\email{nicole@mobiusproject.org}

\author{Emma Lurie}
\affiliation{%
  \institution{University of California, Berkeley}
    \country{USA}
}
\email{emma\_lurie@berkeley.edu}

\author{Deirdre K. Mulligan}
\affiliation{%
  \institution{University of California, Berkeley}
    \country{USA}}
\email{dmulligan@berkeley.edu}

\begin{abstract}
Activists, journalists, and scholars have long raised critical questions about the relationship between diversity, representation, and structural exclusions in data-intensive tools and services. We build on work mapping the emergent landscape of corporate AI ethics to center one outcome of these conversations: the incorporation of diversity and inclusion in corporate AI ethics activities. Using interpretive document analysis and analytic tools from the values in design field, we examine how diversity and inclusion work is articulated in public-facing AI ethics documentation produced by three companies that create application and services layer AI infrastructure: Google, Microsoft, and Salesforce. 

We find that as these documents make diversity and inclusion more tractable to engineers and technical clients, they reveal a drift away from civil rights justifications that resonates with the “managerialization of diversity” by corporations in the mid-1980s. The focus on technical artifacts — such as diverse and inclusive datasets — and the replacement of equity with fairness make ethical work more actionable for everyday practitioners. Yet, they appear divorced from broader DEI initiatives and relevant subject matter experts that could provide needed context to nuanced decisions around how to operationalize these values and new solutions. Finally, diversity and inclusion, as configured by engineering logic, positions firms not as “ethics owners” but as ethics allocators; while these companies claim expertise on AI ethics, the responsibility of defining who diversity and inclusion are meant to protect and where it is relevant is pushed downstream to their customers.
\end{abstract}



\maketitle

\section{Introduction}
Activists, journalists, and scholars have documented ways data intensive practices and systems of surveillance and social sorting produce discriminatory outcomes and representational harms, and connected them to structural exclusion. From evaluations of gender and racial bias in new technologies such as IBM, Microsoft, and Amazon’s facial recognition tools \cite{buolamwini2018gender, snow2018amazon} and search result algorithms for Google images \cite{manjoo2018}, to studies of algorithmic decision-making reproducing or exacerbating structural discrimination in historically inequitable areas such as health \cite{obermeyer2019dissecting}, employment \cite{dastin2018}, and freedom \cite{angwin2016}, these examples shape academic and public discourse around AI ethics. One response to this range of harms has been the rise of AI ethics initiatives in both the public and private sector. In particular, AI ethics initiatives are proliferating in high-profile private sector firms that provide the infrastructure for AI work. These firms have produced a growing set of public facing documents, including but not limited to principles documents \cite{jobin2019global, aiethicsguidelines}. In light of the various discriminatory outcomes data intensive products and services can produce and existing corporate diversity, equity and inclusion initiatives, it is unsurprising that diversity and inclusion appear in public-facing corporate AI ethics documents; although not as first order principles, but often in support of fairness \cite{jobin2019global}. 

Fairness, accountability, and transparency do not exhaust the list of values relevant to data or algorithmic or AI ethics work \cite{fjeld2020principled}. They are part of a constellation of values that overlap, prop up one another, or even conflict with each other. Given the influence these documents may have on the field, understanding how diversity and inclusion are framed, deployed, and made useful in the context of corporate AI ethics documentation merits consideration. 

To gain insight into the work the concepts of diversity and inclusion do in corporate AI ethics, we explore the documents developed by three AI applications and services firms: Google, Microsoft, and Salesforce. We use an interpretive document analysis method to explore ethics documentation -- not as sources for some more or less objective “facts” about data or AI ethics -- but for the insight they provide into how particular actors, institutions, or companies interpret, construct, and make legitimate particular conceptions of diversity and inclusion within ethical AI work.

This paper offers the following contributions: 

\begin{enumerate}
  \item \textbf{Bringing distinct literatures into conversation with each other:} We use conceptual frameworks from the values in design literature to rigorously explore how diversity and inclusion work is being constructed and draw on insights from the institutional law and organizations literature to consider what those interpretations and constructions may mean for the emerging field of AI ethics and related legal fields. 
  \item \textbf{Drifts from civil rights framing:} We document how the concepts and workflows of diversity and inclusion in the AI ethics documents are instrumentalized to fit technical work and drift away from the harms and justifications of civil rights.
\item \textbf{Emergence of Engineering Logic and the Distribution of Responsibility:} We find that firms claim expertise on diversity and inclusion work constructed as amenable to engineering practices while placing the responsibility for ethical deployment on customers.\end{enumerate}

\section{Related Research}
Corporate AI ethics work is ushering in new principles, new workflows, and new professional roles \cite{jackman2016evolving}. Researchers have documented AI-related shifts including: existing corporate subject matter experts taking on new responsibilities to avoid AI-related harms \cite{hirsch2020corporate}; the emergence of new professional roles to “own” ethics in corporate practice \cite{metcalf2019owning, hirsch2020corporate, moss2020ethics}; and the adoption of new institutional structures \cite{newman2020}. The work of individual firms is situated within a broader set of activities aimed at addressing ethics in AI practice and AI-driven systems including the adoption of professional codes of conduct (e.g. ACM,\footnote{\url{https://www.acm.org/code-of-ethics}} DSA,\footnote{\url{https://www.datascienceassn.org/code-of-conduct.html}} Data Science\footnote{\url{https://medium.com/@dpatil/a-code-of-ethics-for-data-science-cda27d1fac1}}); new subfields in computer science and supporting structures such as conferences (e.g. FAccT, MD4SG AIES); new educational initiatives to bring attention to the social, political, legal and ethical implications of technical systems into computer science, and data science courses \cite{fiesler2020we}; funding initiatives to encourage research and educational emphasis on the ethical implications of AI (e.g. NSF FAI\footnote{\url{https://www.nsf.gov/funding/pgm_summ.jsp?pims_id=505651&WT.mc_id=USNSF_41&WT.mc_ev=click}}); new standards (e.g.IEEE Ethically Aligned Design); public policy initiatives attending to the values embedded in algorithmic systems used by government \cite{mulligan2019procurement}; as well as government bans and limitations on the use of specific AI technologies in the policing context \cite{conger2019san}. 

There has been an explosion of ethical AI principles statements--international, national, industry wide, firm-specific, and multi-stakeholder--and analysis of them. These high-level documents shed light on how the field, and distinct actors within it, are discursively framing and theorizing the goals of AI ethics work. While researchers have found significant overlap in the conceptual coverage of AI ethics principles statements, they've noted some discontinuities in the coverage and framing between corporate, and government or multi-stakeholder principles \cite{fjeld2020principled, floridi2019unified, jobin2019global}. Analysis of these principle statements are a useful starting point to explore the work AI ethics is enlisted to do by various actors.

Expanding the set of documents analyzed provides deeper insights into corporate understandings of AI ethics work and their role in operationalizing values from principle statements. These documents are one means by which, as Sara Ahmed puts it, organizations organize commitments \cite{ahmed2007language} and both define and arrange the work of “ethics.'' The documents provide insight into how firms are interpreting key concepts and justifying those interpretations; who is considered responsible for AI ethics work; and how it is implemented. These dimensions are not only significant in understanding the corporate commitments of AI ethics, but also have been found to be contested in  corporate AI ethics principle documents \cite{jobin2019global}. 

In particular, expanding the range of documents past high-level corporate principles sheds light on how firms translate principles into action and provides greater clarity about the problems and solutions they hope to address through AI ethics work. Beyond their literal words, the ways they are written, how they describe things, how the documents work together to produce meaning \cite{drew2006documents} and how they relate to external documents in the emerging field and the documents they in turn reference, deepen our understanding of AI ethics work in the firms. 

In one sense, corporate documents are containers for instructions and value statements; in another, they are a kind of agent, educating clients, the public, and the broader field, articulating and defending values, developing scripts for ethical action that allocate work and responsibility to internal and external actors, and constructing the knowledge and expertise AI ethics work requires. AI ethics is rife with essentially contested concepts that must be translated into local practices to support ethics work \cite{mittelstadt2019principles}. Furthermore, these documents may have explicitly political goals, including influencing, forestalling or preventing regulation \cite{klover2019no}. If that is an aim then early industry action may be key as the adoption of ethical AI practices may be more effective as a proactive than reactive measure  \cite{carberry2012defensive}. %[Carberry & Brayden, 2012 p. 1159] 
 
Viewing corporate AI ethics documentation as part of an effort to establish self-regulation as a preferred path forward foregrounds questions about the moral legitimacy of AI ethics practices; in particular, whether they address the motivating harms and use socially accepted techniques and processes, as well as questions about institutional appropriateness and legitimacy \cite{suchman1995making}. Scholars have raised questions about the ability of companies' ethics efforts to address the ethical wrongs due to conflicts with underlying Silicon Valley logics of ``market fundamentalism, meritocracy, and technological solutionism" \cite{metcalf2019owning}. Scholars have noted how organizational structures and processes adopted by firms shape understandings of legality and compliance through the process of ``legal endogeneity" \cite{edelman2016working}. Thus, the definitions provided and arrangements articulated within these documents influence research agendas and methodological developments, in addition to shaping expectations about the goals of technological development and responsibilities of corporate actors. 
 
In this article, we focus specifically on the ways these documents address ideals of diversity and inclusion. The work of diversity and inclusion in corporate AI ethics documents warrant attention for at least three reasons. 

First, many of the ethical challenges posed by data science and AI revolve around the general problems of bias and discrimination, and particular kinds of allocative, representational, and exclusionary harms \cite{barocas2017engaging}. A wide range of proposed solutions to these problems have been advanced at the institutional and technical level under the labels of “diversity” or “inclusion,” including (but not limited to) strategies for developing more diverse or inclusive sets of training data, algorithmic assessments and audits, diversifying engineering or design teams, and considering a diverse range of customers. Furthermore, the problems motivating the adoption of diversity management programs in the 1980's \cite{dobbin2013origins} and those driving ethical AI initiatives are similarly rooted in concerns with both individual and systemic discrimination. 

Second, because diversity and inclusion have not been identified as core principles in the analysis of AI ethics principles statements they've received less scrutiny. These terms are notably absent from the standard articulations of AI ethics research topics in key conferences solicitations in the field. Given the centrality of diversity and inclusion in general corporate initiatives to address bias and discrimination, their role in the AI ethics field warrants attention. 

Finally, diversity, equity, and inclusion discourse has been identified as a site of contestation within data ethics and related fields. These terms while distinct, tend to travel together: diversity is focused on the representation of different group affiliations of cultural significance, inclusion is centered on members of groups being included in key decision-making, and equity is the absence of systemic disparities across distinct groups \cite{bernstein2020diversity}. 

Ruha Benjamin has flagged diversity and inclusion as a kind of “happy talk” that acknowledges cultural difference without challenging structural inequalities \cite{benjamin2019race}. %(p. 148).
Sareeta Amrute notes that the corporate embrace of diversity at once makes things like racial diversity visible and celebrated, while also underplaying or obscuring the broader historical and political dynamics that inform systems of racial difference \cite{amrute2020bored}. Along a similar vein, Anna Lauren Hoffman discusses how inclusion in data technologies, if not accompanied by shifts in power, can further vulnerability to  “data violence” by normalizing otherwise oppressive structural conditions \cite{hoffmann2020terms}. As Devin Guillory points out in his discussion of anti-Blackness in AI communities, diversity is often valued only insofar as it is seen as improving corporate performance—a “predatory view…in which the worth of underrepresented people is tied to their value add to in-group members” \cite{guillory2020combating}. In this way, Anna Lauren Hoffmann claims, diversity and inclusion in data science and technology largely represent “an ethics of social change that does not upset the social order” \cite{hoffmann2020terms}.

%(p. 12). 

This aligns with law and organization scholarship where Lauren B. Edelman, Sally Riggs Fuller, and Iona Mara-Drita \cite{edelman2001diversity} found that diversity management rhetoric diluted the aims of civil rights law by expanding beyond legally protected categories to include thought, lifestyle, culture, dress, geography among others, thus distancing diversity from the goals and logics of civil rights laws, and aligning it with corporate success rather than ameliorating specific discriminatory harms. More recently, Ellen Berrery documents a shift away from civil rights in corporate statements around diversity which now talk about diversity in business products, and position diversity as instrumental in global business expansion and success rather than responsive to histories of oppression and exclusion \cite{berrey2015enigma}. Edelman argues that “(t)he rhetorical transformation from civil rights to diversity” replaced “the public commitment to minority hiring, and in particular to the hiring of black Americans…” with a “commitment to diversity--a construct that is almost universally accepted as valuable and yet does little to promote race and gender equality in the workplace” \cite{edelman2016working}.

\section{Methods}

We selected three companies--Microsoft, Google, and Salesforce--as our sites of study as they provide application and service layer infrastructure \cite{wirtz2019integrated} to customers seeking to do AI and data science work. These three companies produce a wide range of documents related to AI ethics including ethics guidelines, blog posts, educational modules, product documentation, and tooling for operationalizing ethics-related values. They rank among the most influential technology firms globally. Combined, their enterprise cloud services constitute a large portion of the market for enterprise cloud and data analytics services.\footnote{\url{https://kinsta.com/blog/cloud-market-share/}} These firms are highly engaged with regulators and policymakers considering whether and how to regulate AI to protect public values, Given the ongoing ambiguity of the external policy environment, the discursive practices and organizational choices of these influential players are primed to shape the field as other firms seeking to signal legitimate and responsible behavior to external stakeholders follow their example--a process of mimetic isomorphism \cite{dimaggio1983iron}. These firms are also shaping the research agenda of ethical AI through publications, participation in scholarly communities, and financial support. The numerous sites of interaction between corporate AI ethics professionals and the broader emerging AI ethics field, and the relative wealth of AI ethics approaches emerging from these firms--both their practice and research groups--create an environment where corporate practitioners are likely to play an influential role in diffusing concepts, policies and practices across firms and into the broader field \cite{bamberger2015privacy}. This is not to say that these companies can wholly determine understandings and practices -- their own or others -- but rather to acknowledge the relative potential influence of these documents on the emerging field of ethical AI. We initially sought to include Amazon in our corpus, but a lack of documents about AI ethics made it infeasible to include the company in our analysis.

Our document collection, coding, and analysis proceeded as follows. We first collected a set of documents from Microsoft, Salesforce, and Google that related to values and ethics in AI by searching relevant terms on Google’s search engine and pulling links from the first page of results. We added to our set of documents through purposive sampling of additional documents from these three companies that were hyperlinked in the documents that emerged from the Google search. Our set of documents include ethics guidelines, blog posts, educational modules, documentation related to products, and tooling for operationalizing ethics-related values. Because research divisions are sometimes siloed from the company’s main operations, we did not include articles from research divisions unless they were focused on “tooling” -- in other words, focused on operationalizing ethics-related values in the product development process. For this analysis, we then identified a subgroup of 46 documents that reference diversity, equity, and inclusion, including references to civil rights law (ex. protected categories) and human rights.

\begin{figure}
   \centering
   \begin{subfigure}[b]{0.5\textwidth}
     \centering
     \includegraphics[width=\textwidth]{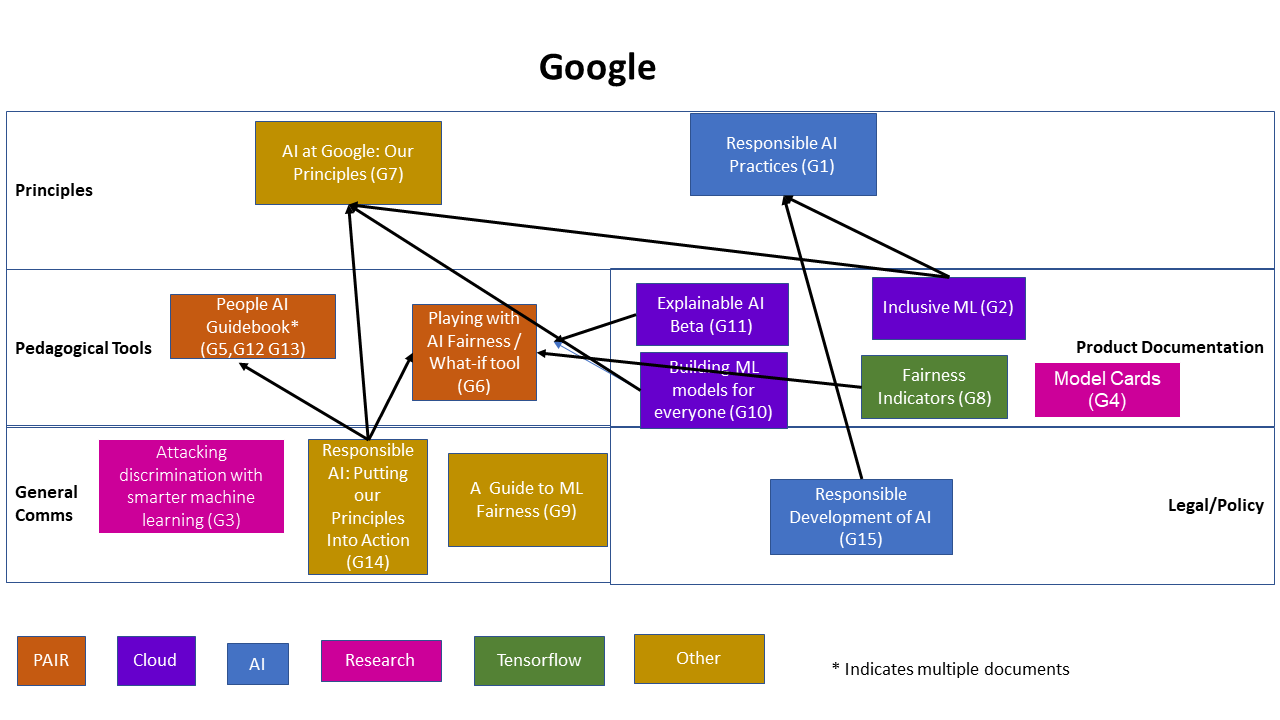}
     \label{fig:google}
   \end{subfigure}
   \hfill
   \begin{subfigure}[b]{0.5\textwidth}
     \centering
     \includegraphics[width=\textwidth]{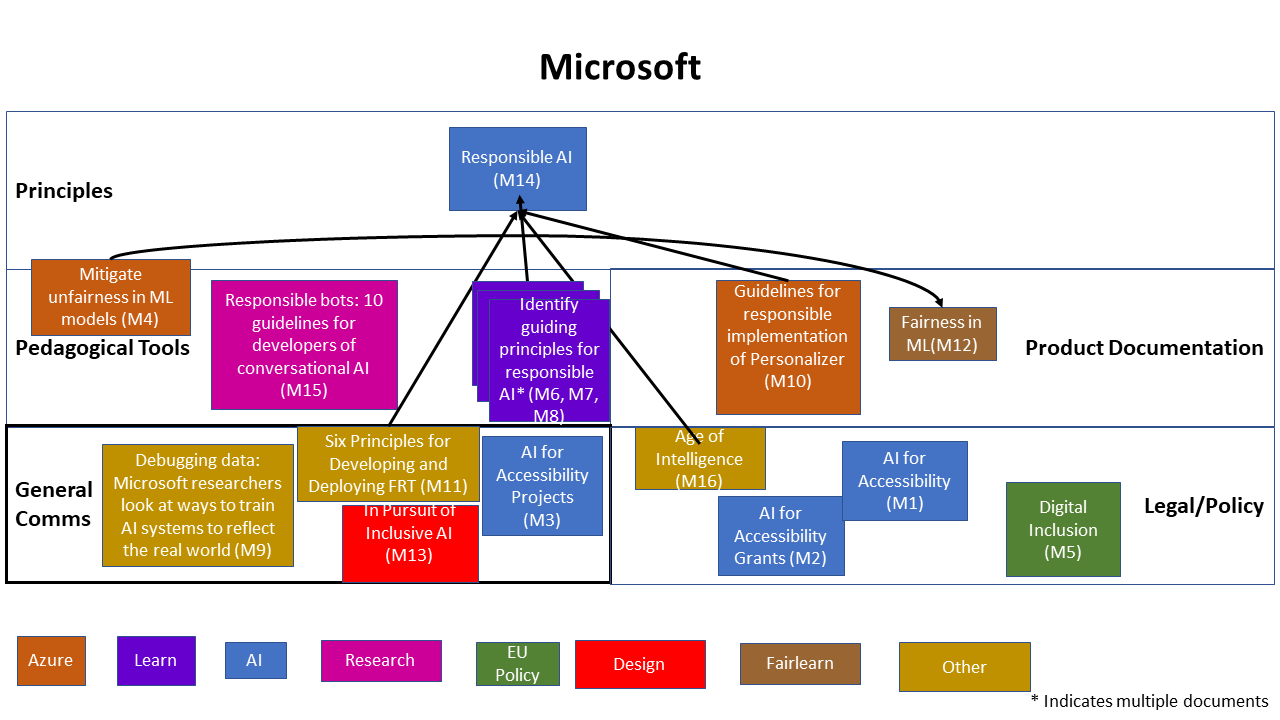}
     \label{fig:microsoft}
   \end{subfigure}
   \hfill
   \begin{subfigure}[b]{0.5\textwidth}
     \centering
     \includegraphics[width=\textwidth]{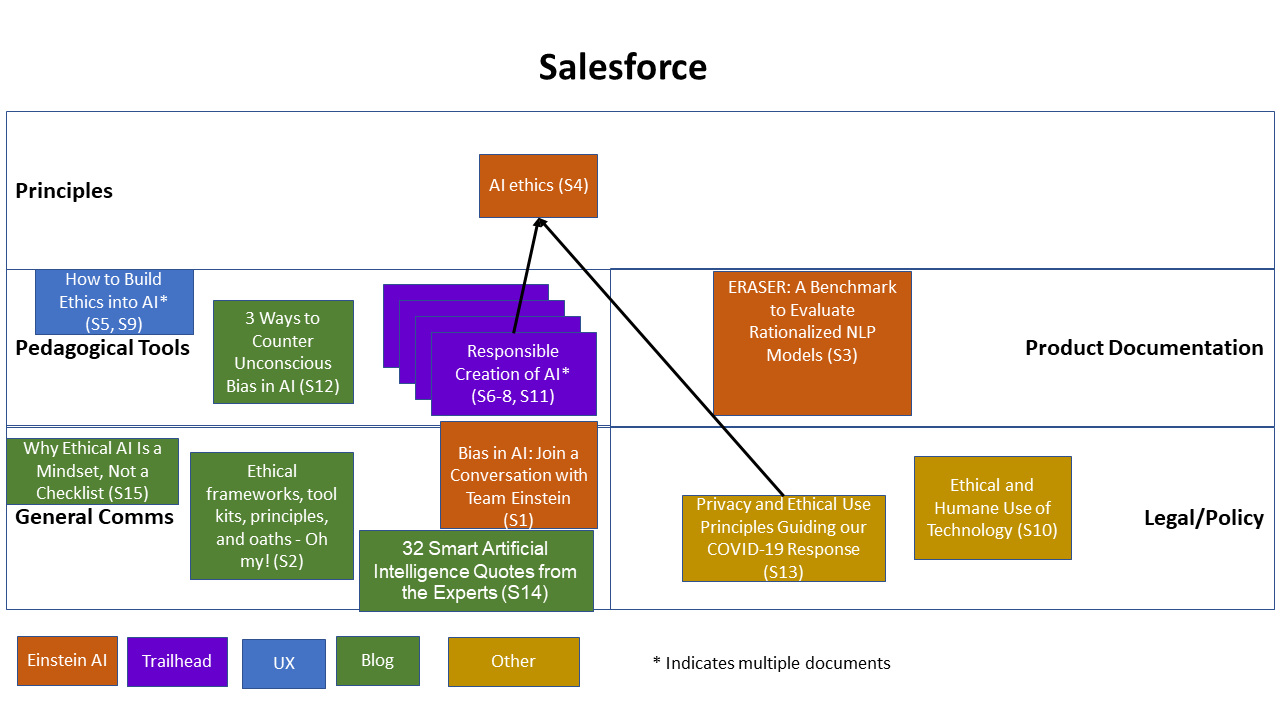}
     \label{fig:salesforce}
   \end{subfigure}
    \caption{A visualization of the corpus. We deconstructed the corpus into 5 categories: 1) principle documents, 2) pedagogical tooling, 3)product documentation, 4) legal or policy documents, and 5) general communication documents. The documents are color coded to illustrate the various internal teams that produced the documents. Google, Microsoft, and Salesforce have varying distributions of the corpus and different levels of connectedness between documents.}
    \label{fig:mappings}
\end{figure}

As illustrated in the Figure~\ref{fig:mappings}, our document collection strategy surfaced documents from a variety of departments that serve different purposes. We code the documents to illustrate the various departments documents originate from and divide the figure into sections to represent the different document functions. In the pedagogical tools sections, we include professional development tutorials, such as Salesforce Trailhead (S6-8, S11)\footnote{See the supplementary materials for the full list of documents.}, as well as guidelines that are not attached to a particular product, such as Google's People + AI Guidebook (G12-13). The product documentation category includes documents that describe new products created to address AI ethics (M12) or where an application of the technology involves a discussion of AI ethics (G11). The Legal/Policy category includes documents that come from the legal or policy departments (S13) or that are external funding initiatives that further policy goals (M1-3). The General Comms category holds the broadest range of documents, ranging from interviews with engineers (e.g. M9) to blog posts that act as company op-eds (S15). 

Our analysis is guided by the methods of critical discourse analysis and critical informatics. Broadly, discourse refers to the use of language relative to social, political, and other formations \cite{jaworski1999discourse}.  Critical discourse analysis is particularly appropriate given that certain actors are better positioned to shape the meanings that get attached to technology than others \cite{van1996discourse}. In Foucauldian terms, discourse as a term is used more specifically to describe the (historically-bound) rules and assumptions that make knowledge possible. It simultaneously reflects and shapes social orders and practices, informing both what people do in practice and how people understand and represent those practices at any given time \cite{van2008discourse}. In this way, discourse is not reducible to a practice or a representation; instead, discourse is a kind of mediating layer that enables the successful accomplishment of either. If discourse is defined as the recontextualization of social practices, and knowledge is produced at one level and then embedded into another, where it is made to serve contextually-defined goals or purposes \cite{van2008discourse}, analyses of discourse take up and scrutinize representations in the form of texts as sites where discursive rules and assumptions are encoded.

 To understand how the contested concepts of diversity and inclusion are specified and enacted in these corporate documents we draw on methods and analytic tools from the values in design field. Values in design research places technical artifacts, as well as human action and policies, within the scope of ethical analysis. It encompasses both research analyzing how values are enacted in existing sociotechnical systems and constructive work aimed at bringing selected values into design processes. While the objects of study are sociotechnical systems, where to draw the boundaries of such a system, and where and how to study values in relation to it are ambiguous and contested \cite{shilton2014see}. Similarly, the contested and contextual nature of values such as privacy, fairness, transparency, etc. complicates efforts to study values in existing sociotechnical systems, and importantly for the work of AI ethics, to build to support them. To address these challenges, researchers have developed frameworks that clarify various ways to see and position values in relation to sociotechnical systems \cite{shilton2014see,mulligan2020concept}, and analytic tools to unpack and model contested values \cite{flanagan2005values, mulligan2020concept, mulligan2019thing}.

\section{Findings}

\subsection{Configuring Diversity and Inclusion}
The documents configure diversity and inclusion and the work associated with them in particular ways:  the term equity is largely absent, and at times fairness is substituted; diversity and inclusion are expanded to include a wide range of attributes, as they are in managerial diversity rhetoric, but attach to technical artifacts in addition to people; technical experts and technical methods and tools are the means of accomplishing AI ethics goals. 

\subsubsection{The disappearance of equity and the rise of fairness:} The standard triad in corporate management--diversity equity and inclusion--becomes a diad. While diversity and inclusion appear frequently, and together, equity appears relatively infrequently and separately. Where we might expect to see the word equity referenced as a justification for diversity and inclusion efforts these documents selectively substitute in fairness. Equity is a form of fairness; however, fairness can take other forms such as procedural regularity or equal treatment. 

This Google Cloud document (G10) provides a clear example of the substitution of equity for fairness: 
\begin{quote} 
\textit{Fairness} is the process of understanding bias introduced by your data, and ensuring your model provides \textit{equitable} predictions across all demographic groups. Rather than thinking of fairness as a separate initiative, it’s important to apply fairness analysis throughout your entire ML process, making sure to continuously reevaluate your models from the perspective of \textit{fairness and inclusion}. [emphasis added]
\end{quote}

\subsubsection{Diversity and inclusion in action:} The documents provide a strong sense of the companies’ opinions of what entities are responsible for diversity and inclusion work, as well as which professionals and which tools are necessary to execute on that responsibility. They also provide a general, if somewhat more ambiguous understanding of what this work aims to protect or protect against. 

To unpack how the problems and solutions related to diversity and inclusion work are framed in the documents we draw on a conceptual analytic created to reflexively work with the contested value of fairness \cite{mulligan2019thing}. We use the analytic’s meta-dimensions of protection and provision, and the sub-dimensions within them, to explore how the documents frame what diversity and inclusion work: ought to provide and whom it ought to protect (the problem to be solved/end state to be achieved); and the who and how of implementation (solutions). While the fairness analytic was designed specifically to facilitate interdisciplinary research and practice with the value of fairness we found these meta-dimensions equally useful for exploring the values of diversity and inclusion in this corpus.\footnote{See the relevant sections of the fairness analytic in the supplementary materials or in \cite{mulligan2019thing}.}

Interrogating the dimension of protection requires identifying both the \textit{target} of protection and the \textit{subject} of protection. To identify the target, ``the ideal end state toward which fairness
aspires," the analytic suggests asking, ``What should fairness provide?" To identify the subject, the actors or entities for whom fairness is provided, the analytic suggests asking, ``Fairness is at stake for whom or what?''

Interrogating the dimension of provision requires identifying the \textit{provider}, \textit{mechanism}, and \textit{implementer}. To identify the provider, the actor charged with being fair, the analytic suggests asking, "who or what is supposed to behave fairly or avoid unfair behavior?" To identify the mechanism, the modalities used to support fairness, the analytic suggests asking, "how is fairness operationalized?'' To identify, the implementer, the actors tasked with operationalizing fairness through chosen modalities, the analytic suggests asking, "Who brings fairness into practice?"

\subsection{Dimension of Protection}
The companies present themselves as experts on ethical AI, but they position customers using their AI applications and services as responsible for ethical work and outcomes. This distribution of responsibility leaves the \textit{target} and \textit{subject} of diversity and inclusion to be determined by the customer and therefore both vague and ambiguous within these documents. 

The companies explain the various aspects of the technical product development lifecycle where diversity may be relevant, including assembling a diverse team, using diverse data to train your model, and testing on diverse users; however the customer is responsible for figuring out which are important to a specific project. Similarly, the companies explain that diversity can be relevant to an expansive range of humans and technical artifacts. But here too, determining which humans or technical artifacts must be diverse to advance specific goals is left to the customer's discretion. Where recommendations are provided, they tend to be abstract rather than concrete examples of the diversity that would serve to make a specific product or service fair. For example, Microsoft discusses how in hiring, school admissions, and lending, an AI system could withhold opportunities for “certain groups” or be “much better at picking good candidates among a specific group of people than among other groups” (M4), while Salesforce uses the ambiguous term “values-based bias” (S9). All three companies use terms like “sensitive features” (M12), “sensitive characteristics" (G7), or “sensitive fields” (G7) as containers to allow customers to designate whatever traits they deem worthy of diversity, fairness, and inclusion work.  

These generic terms require customers to determine who the \textit{subject} of diversity and inclusion is, which allows the companies tools and services to support system design across a range of applications and context. At times they are clear about these terms being substituted for a narrower set of legally relevant attributes, and even justify this substitution by discussing how AI ethics can offer broader protections than law. Microsoft deliberately avoids use of “protected attributes” from anti-discrimination law explaining that they "seek to apply group fairness in a wider range of settings" (M12). In a document about their Fairlearn product, they ask engineers to define sensitive features by asking themselves which groups of individuals are at risk for experiencing harms, rather than citing specific protected categories (M4). Likewise, Salesforce cautions that even if AI models are making decisions where it is legal to rely on these protected characteristics, AI ethics must sometimes extend beyond law because “it still may not be ethical to allow those kinds of biases” (S8).

\subsection{Dimension of Provision}
This focus on the customers’ goals is supported by a shift in the people and tools of implementation. The customer is the \textit{provider} responsible for attending to diversity and inclusion. Engineers and designers are the \textit{implementers} tasked with using repurposed and augmented design and engineering methods--\textit{mechanisms}--to support diversity, inclusion and fairness work. 

The tools (G8), guidelines (M10, S9), and methods (S8) described in the documents, frame diversity and inclusion as the work of engineers, designers and product managers. Many of the interventions are aimed at the system design itself such as using diverse data sets, diverse test users, and diverse use cases, as well as specific technical mechanisms for modeling constructs of fairness or exploring biases in inputs and outputs. Diversity is placed within the domain of engineering, recentered on product optimization and supported through new processes of diversifying datasets and labelers, designing for diverse contexts, and evaluating features through diverse user testing. While methods include managerial approaches such as using diverse technical teams, they are nonetheless interventions in the design and engineering processes. To support this reorientation, diversity and inclusion becomes a potentially important consideration for the particular teams developing a product (rather than the workforce as a whole), the user testing cohort, or customer education. 

The focus on engineering work expands the subjects of diversity and inclusiveness to non-humans. For example, the documents discuss various ways in which system inputs--such as datasets (G8, G2, S4), and use cases (G14)--can be diverse and inclusive. This extension of the subjects of diversity reflects the needs of AI and ML engineers and practitioners who work with data and may rarely interact with the humans affected by their systems.  

The documents set out a wide range of attributes along which diversity can be considered across human and technical subjects. In the limited instances where a company references anti-discrimination law, they mention legal categories such as sex, race, and class. However, diversity often drifts away from categories based on histories of disadvantage and oppression toward a broader set of dimensions. For example, Salesforce instructs: “Development teams should strive toward diversity in every area, from age and race to culture, education, and ability” (S6). Similarly, Microsoft discusses including team members that have “different backgrounds, experiences, education and perspectives” and urges clients to consider AI’s impacts on “certain groups” (M7). 

Within pedagogical tools and product documentation legally relevant attributes are replaced with traits more amenable to technical forms of observation and measurement. For example, in Google’s Fairness Indicators, they state: 
\begin{quote}
“It is ill-advised to build an image classifier for race or ethnicity, because these are not visual traits that can be defined in an image...instead, building a classifier for skin tone may be a more appropriate way to label and evaluate an image” (G8).
\end{quote}

Similarly, in Microsoft's Personalizer tool, user demographics such as race, gender, and age are described as legally troublesome to consider, whereas behavioral data collected by technology products is framed as a more useful definition of identity. The document's section on inclusiveness makes no mention of diversity, discrimination, or access to opportunity, instead focusing on how personalization can support inclusion asking engineers to:
\begin{quote}``...question how relevant demographic information is when you have actual interaction, contextual, and historical data that relates more closely to the preferences and identities of users" (M10).
\end{quote}
Shifting the focus to observable behavioral data aligns inclusion with the engineering work that this document is meant to support; Personalizer is a tool built to help adjust application behavior to context.

Finally, to further support engineering work, all three companies provide various technical definitions and metrics of fairness. Group fairness and demographic parity are mentioned multiple times across companies (M4, S8, G6) as a way to potentially test to ensure fairness in diverse datasets. The tradeoffs of metrics like false positives and false negatives are discussed in the context of evaluating inclusive and diverse models (G6, G8, M12, S7). The documents reference academic work discussing the multiplicity of types of fairness (S6). These technical definitions and metrics become part of the language of ethical and inclusive AI.

Making the concepts legible to technical practitioners, \textit{instrumentalization}, makes diversity and inclusion more amenable to engineering mechanisms, however it also constructs and bounds the problems and solutions of AI ethics. First, engineering and design failures with particular ethical impact due to culture, history and context are are transformed into more generic classes of technical errors and primed for technical solutions. 

For example, in Google’s People + AI Guidebook a case study of users in a region consistently rejecting meat based suggestions during a “certain time of year" due to religious obligations, is portrayed as a generic “context error” that can be fixed through better personalization and in particular the collection and use of location data (G12). Tools that help users visualize and analyze ML models without writing code which help tech teams “build interpretable and inclusive AI systems from the ground up” (G11) are enlisted in ethics work. Tools that allow customers to build custom machine learning models are marketed as inclusive because they can help development teams produce “contextually relevant ML systems” (G2).

The personalization supported by machine learning is cast as integral to solutions at scale. In reference to its tool, Personalizer, Microsoft discusses how inclusiveness means providing “personalized experiences for accessibility-enabled interfaces” and adjusting “application behavior to context” (M10). Incorporating "context" is not only part of the mechanism of inclusive AI, but becomes part of the branding of inclusive AI as a better product.

These reframings of diversity and inclusion in terms tractable to technical work allows the companies to position the very tools and services that ethical AI efforts are meant to interrogate as the solutions. For example, tools that teach technologists to “remove exclusion” with techniques such as corpus-level constraints or debiasing word embeddings (S11) become part of the ethical AI toolkit. Technical work such as assuring the use of diverse data sets, and managerial strategies such as using diverse teams, are positioned as tools that can address the potential harms of technology (AI reinforcing stereotypes), and yield better products (helping AI systems make better inferences based on context). While the documents also discuss non-technical solutions, they consistently loop back to technical fixes, urging clients to identify whether users’ values or cultural considerations are being overwritten, and emphasizing “modify(ing) the training data, labels, and algorithms to represent the diversity of values” as the solution to preventing bias (S11). The problems posed by data scientific work can be largely, though not completely, solved by engineers and designers using the advice, tools, and methods offered by the firms. 

\section{Discussion: diversity and inclusion reconfigured} 

The AI ethics documents resonate with the drift away from civil rights found in corporate DEI documents \cite{berrey2015enigma}, but chart new ways of enacting diversity and inclusion work. The expansive forms of diversity considered relevant, the new technical artifacts to which diversity can attach, and waning connection to equity attenuates the connection between diversity and inclusion work and the specific harms that inspired it and the logics and discourse of the relevant legal fields. The emphasis on technical experts, processes and solutions, however, offers a distinct perspective on the professionals and workflows viewed as central to diversity and inclusion work. This orientation toward technical work and professionals shapes AI ethics work in two ways. First, it centers technical artifacts rather than business models as the sites of intervention and technical interventions rather than organizational or regulatory, for example, as the solutions. Second, the framing ushers in engineering logic that positions these firms as experts on Ethical AI work but places responsibility for determining the goals and outcomes to be supported by diversity and inclusion work on customers. With this logic, firms and engineers are not responsible for the real world ethical outcomes to which their expertise and tools contribute.    

\subsection{Drift}
The documents reveal a vision of diversity that, like the diversity rhetoric arising in the management literature beginning in the mid-1980’s, drifts away from civil rights justifications and the specific forms of discrimination it addresses. As in the “managerialization of diversity” the documents expand the categories of desirable diversity to include thought, lifestyle, culture, dress, and geography, and rarely associate diversity with protecting civil rights \cite{edelman2001diversity}. Diversity is positioned as instrumental to achieving product or business objective, such as products that serve diverse markets. Inclusiveness is viewed as instrumental to democratizing AI and building customer trust, and producing better business insights. The lack of specificity and clarity about the purpose and justification of diversity and inclusion work, and deference to customer preferences, further distances ethical AI work from the equitable aims of civil and human rights laws, and, as we discuss below, places it largely outside the firms’ responsibility. 

But the drift takes new forms. First, technical artifacts, not just humans, become a locus of diversity and inclusion work. The new emphasis on diversity and inclusion of non-human subjects -- datasets, use cases, etc. -- is surely necessary to advance equitable outcomes in AI systems. Yet at the same time, firms can enact these diversity and inclusion practices--producing better products or improving market access--without addressing the distinct disadvantages minoritized groups have experienced throughout history that, for example, produce the biased data sets, or addressing internal practices of exclusion or oppression. 

Second, equity, a particular form of fairness, is replaced by a multiplicity of potential fairnesses. Here too, this broadening may be necessary to support technical work that attends to the various forms fairness can take in different contexts \cite{mulligan2019thing,chouldechova2017fair}. The move to fairness allows designers to tap into the plethora of narrower technical formalizations, framings, and metrics produced by the technical research community (G6, G7). Yet substituting fairness for equity does political work. It replaces a normative commitment to equity in diversity and inclusion work with an undifferentiated commitment to supporting various forms of fairness. As Mittelstadt notes, "At best this conceptual ambiguity allows for context-sensitive specification of ethical requirements for AI. At worst, it masks fundamental, principled disagreement and drives AI ethics towards moral relativism" \cite{mittelstadt2019principles}. This is illustrated in Google’s “Playing With AI Fairness” document (G6) that presents all definitions of fairness as being equally valid. “As the morning breaks, the five experts are still collegially arguing...Which sense of fairness is the fairest in the land? There is no right answer…” 

While we cannot fully understand the potential implications of the drift in diversity and inclusivity work these documents suggest, it highlights two critical lines of inquiry.

At this moment, corporations enjoy substantial lee-way to define what AI ethics requires of them, including what expertise is required and who possesses it. Consistent with the positioning of diversity and inclusion work within these documents, individuals working on ethical AI within firms position it as distinct from the law and distanced from compliance \cite{hirsch2020corporate,metcalf2019owning}. Yet, regardless of the disconnections from law in these documents or in corporate work flows, AI ethics work is entangled with law. The ideals of inclusive workplaces and services, respect for diverse individuals along attributes of race, ethnicity, ability, and gender, and fairness in treatment and outcomes of individuals are the heart of civil rights laws. The motivation for AI ethics work arises, at least in part, from high profile examples revealing the role algorithmic systems can play in perpetuating systemic harms, masking disparate treatment, and producing discriminatory outputs. The examples used in these documents highlight concerns with biased performance along racial and gender lines: Salesforce links to articles about Google Photos labeling Black faces as "gorillas" and gender bias in Google Translate (S3, S9), while Google references soap dispensers that do not recognize dark skin tones (G2). While specific laws are rarely mentioned; law is implicitly acknowledged through general examples about hiring and lending, and the COMPAS system \cite{angwin2016}. 

Thus, while positioned as distinct from compliance, the overlap with the substantive harms civil rights laws seek to address all but assures that these firms’ approaches will shape the law. The extent of such influence is uncertain, however the lack of clarity and guidance coming from external actors suggests it could be substantial. First, where the expectations of external stakeholders are unclear, firms mimic the structures of their peers in an effort to signal legitimacy and rationality, a process of “institutional isomorphism” \cite{dimaggio1983iron}. Second, as firms adopt policies, processes, and structures to symbolize attention to the ethics of AI, those structures may become equated with responsible action and shape understandings of compliance, and over time regulators and other legal actors may take their mere existence as evidence of legality \cite{edelman2001diversity}. This entanglement with the law \cite{edelman2016working} raises questions about the potential implications of the reconfigurations of diversity and inclusion in these corporate documents for the firms'--and over time, legal actors’ and institutions’--understanding of what civil rights law requires of those who develop and deploy AI systems. Is it necessary, appropriate, sufficient to substitute terms tractable to technical systems (skin tone, geographic regions) for legally relevant concepts (race and religion)? Will legal institutions expect corporate AI development work to use diverse teams or testers? While these AI ethics documents rarely speak of rights or reference legal frameworks, the influence they may have on protections for civil rights in the development and use of AI systems warrant attention to the risks and benefits they pose. For these reasons, the corporate configurations of ethical work--what it means, who does it, and who is responsible for it--deserve scrutiny. 

Surely some of this reconfiguration is necessary, and potentially beneficial, as it enables technical work to contribute to products and services that yield more equitable or fairer outcomes. It brings technical practitioners to the table in two distinct ways. First, the abstraction and formalization, and emphasis on mechanisms rather than policy found in these documents frames diversity and inclusion work as part of technical practice. This framing may encourage engineers to view ethics as, at least in part, a legitimate, potentially even mandatory, aspect of practice. Similarly, the emphasis on design processes as sites for diversity and inclusion efforts creates space for UX and other design professionals to engage in values work. Overall this aligns with moves to pragmatic engagement with ethics during design that position engineers and designers as essential participants in surfacing and protecting values \cite{friedman2002value, flanagan2005values}. The guidance on technical approaches and considerations, while still ambiguous, provide some connection between the firms' principles and the construction and use of their tools and services. It begins to provide some sorely needed pragmatic direction to guide ethical work on the ground. 

Privacy research underscores the importance of bringing engineers and designers into values work. Scholars and regulators emphasize the importance of embedding privacy into technology design and business processes \cite{bennett2017governance, regulation2016regulation, regulation2018ccpa}. Research on implementing privacy in the organizational setting notes the importance of integrating privacy into the regular workflow of business units and technical design so that it can “engage employees at times and in venues where privacy concerns can influence technical systems and business practices” and be more responsive to “contextualized understandings of privacy, and privacy harms” \cite{bamberger2015privacy}.%[Bamberger and Mulligan 2015. p. 180]. 

In addition to making ethical work actionable by technical practitioners, the drift may do political work within the firms by establishing a boundary that creates space for technical practice. Researchers have found that corporate AI ethics work exists in the interstices of other corporate activity, is distinct from “corporate roles that...demonstrate compliance with laws (business ethics, compliance and whistleblowing)” and that “ethics owners...facilitate compliance with informal and evolving standards, prepare for potential future regulations, and attempt to prevent social harm” \cite{metcalf2019owning}. Internal ``ethics owner'' suggest that placing ethics outside compliance can make ``'doing ethics' logistically easier” \cite{metcalf2019owning}. %[Metcalf et al p.458]
If AI ethics work is viewed by those inside the firm as “...venturing beyond...law and into the realm of substantive value choices...”, then firms may want to limit the risk that this work comes to inform external perspectives on compliance. Through our immersion in the field, we are aware of struggles within some firms over what AI ethics ought to do within firms, and how it relates to ongoing work streams around privacy, human rights, accessibility, research ethics, etc. Terminology is one way to assert a boundary. This boundary may create room for beyond compliance activities.

The substitution of fairness for equity, and emphasis on technical formulations of it, may be particularly important in carving out space for technical practitioners. Use of the term fairness may allow divergence between the AI ethics work and other values-related work that is more directed by regulatory concerns and compliance. Portraying AI ethics diversity and inclusion work as separate from firm DEI efforts and legal compliance, may create new opportunities and license for technical employees to engage in ethical inquiry. This may open space for new “ethics advocates” who hold some authority and can “lobb[y] for, social and ethical concerns within the design process" \cite{shilton2014see}. Above, we noted the potential for evolving corporate understandings and practices in the AI ethics field to shape understandings of what civil rights law demands of AI related work and systems. Here, we note that the substitution of fairness for equity may be an intentional effort to limit the influence of AI ethics work on understandings of compliance. 

Finally, by legitimating the work of technical practitioners, the framing may also reduce the emotional labor of “work that seeks to address broader politics and harms” \cite{wong2020thesis}. By centering the work of technical practitioners this framing of AI ethics work may thus provide the means for technical practice, create space for it, and reduce the perceived and actual risk of bringing ethics into technical work. 

While fostering and legitimating technical values work is important, research notes the importance of maintaining the salience of social and political values at a strategic level where it is often in tension with, rather than in support of, corporate interests and logics. The need for values to be embedded and perceived to be part of the mundane day-to-day firm work, yet also operate at the political level poses implementation challenges \cite{bamberger2015privacy}. To the extent limiting discrimination or advancing an anti-subordination agenda requires technical work, which it surely must, then efforts to make it legible and tractable in the work of engineers and designers--the selection of data sets and testers, etc.--are necessary. The important question is whether the emerging workflows required to bring diversity and inclusion work into technical practice are meaningfully connected, and supported through institutional structures that keep the broader public interest in a fair and inclusive society--“help[ing to] eliminate relationships of domination between groups and people based on differences of power, wealth, or knowledge” and “produc[ing] social and economic benefits for all by reducing social inequalities and vulnerabilities”--central to the work of the firm and its customers \cite{montrealethics}.

The documents we reviewed provide limited insight into the answer to this question. However, we note that they do not reference broader corporate diversity, equity, and inclusion (DEI) initiatives or the human rights practices with these firms, nor do they explicitly reference frameworks or legal institutions associated with advancing and protecting civil rights. Salesforce documents intermittently mention their overarching corporate principle of equality (e.g. S5) yet, the more actionable and pedagogical documents rarely mention them. Absent such connections, technical work risks losing its political mooring and meaning \cite{bamberger2015privacy, mcgaley2006critical}. Importantly, researchers exploring technical work with the value of privacy find that it "requires a thorough understanding of the context: a holistic analysis of the risks and threats in that given context; an ability to systematically analyze those risks and threats; while reconciling the privacy and functional requirements using state of the art research results" and that this requires practitioners to be informed about not only "the state-of-the-art research in security and privacy technologies" but "legal frameworks and the current privacy and surveillance discourses" \cite{gurses2011engineering}. A lack of concrete connections between engineering practices and the expertise and approaches of other subject matter experts within the firm could limit the efficacy of technical practitioners' attempts to operationalize values such as diversity, fairness, and inclusion.

Framing diversity and inclusion AI ethics work as the domain of technical experts and solutions occludes other sites of ethical intervention such as business models, and solutions such as the adoption of regulatory frameworks \cite{greene2019better}. The documents do not advocate processes that connect technical practitioners with other experts within the customer's organization who may be aware of context related diversity, inclusion, and fairness concepts captured in law or other normative sources. Thus, while these documents bring engineers and designers to the ethics table, it also sets them apart. In doing so they fail to connect engineers with other professionals whose expertise might both help engineers make more nuanced decisions at the technical level and open up new solutions. If "ethics continues to be seen as something to implement rather than something to design organizations around, “doing ethics” may become a performance of procedure rather than an enactment of responsible values" \cite{metcalf2019owning}. The capacity to do ethics work focused on data sets and use cases and various fairness metrics without advancing equity poses a risk of this sort of performance.

Milena Doytcheva connects corporate diversity policies in France that jettison race and ethnicity in favor of ‘universalisation’ to the adoption of “racist structures and power relations within purportedly race-conscious procedures.” Importantly for our discussion of engineering logic below, Doytcheva relates this “unversilization” to the outsourcing of diversity standards and measurement to technocratic self-regulatory instruments, bodies and consultants \cite{doytcheva2020governing}. The focus on technical approaches and metrics within these documents and lack of clear connection to civil rights aims may pose similar risks here.

\subsection{Engineering Logic} 
These documents portray diversity and inclusion work as primarily technical work that centers the expertise of engineers, designers, and product managers. This technical orientation brings engineering logic into diversity and inclusion work. The logics of engineering, including commitments to abstraction, formalization, and designing mechanisms--not policy--shape these documents and contribute to the orientation to different kinds of human roles and technical artifacts. The concepts of diversity, inclusion, and fairness are formalized and abstracted, linked to an emerging technical literature, and readied for deployment by customers across a wide range of contexts. Technical processes and solutions are retooled or rebranded as part of ethical practice.  Just as Edelman et. al found law being managerialized through diversity rhetoric, we see a new phenomena where problems voiced in legal and political terms--bias and discrimination, race and gender--are replaced and augmented with technical definitions and practices, which we call \textit{instrumentalization}. %, instrumentalizing concepts historically tied to ethical and legal ideals. 

Engineering logic drives particular arrangements of expertise and responsibility. In these documents, firms position themselves not as expert “ethics owners” but \textit{ethics' allocators}. Consistent with other scholarship documenting AI practitioners' belief that ethical responsibility for AI systems should be distributed across a range of actors \cite{orr2020attributions}, these documents reveal firms distributing much of the responsibility for ethical decisions and outcomes to customers. The companies define and claim expertise over the instrumentality and reasoning that customers need to solve ethical AI challenges. The educational style of many of the documents--both pedagogical tools and product documentation, as well as those in the general communication category--convey the expert role of the firms. Companies showcase their ability to explain the complicated, contextual, and indeterminate nature of ethics work which helps them shore up both their position as experts and their normative stance that the responsibility for sorting out ethical obligations is customer work informed by whatever regulations or other forces constrain them. 

Allocating the responsibility for ethics to customers is consistent with engineerings’ professional identity reflected in efforts to distinguish between mechanisms and policy, and technology design and policy outcomes. It also responds to a growing set of concerns about the risks that abstraction and solutionism pose to meaningful ethical work and outcomes \cite{selbst2019fairness}. The allocation of responsibility for figuring out what ethics requires to customers suggests that the unit of analysis is the sociotechnical system, rather than the underlying technical components. %, but rarely provide concrete non-technical strategies other than diverse data and teams. 

The documents position ethics as contested and contextual, ethical outcomes as contingent on a multitude of user choices, and importantly as something to be assessed at the sociotechnical rather than solely technical level. Through questioning, the AI ethics documents attempt to surface values for reflection and attention, while maintaining responsibility for enacting values with the customer. The question format found in these documents aligns with a plural understanding of values, and the recognition that designers “make sense of values not at remove... but in the often-confused design situations in which a value has value” and the role that reflection and questioning play in sorting out "the value values have in context" \cite{Jafarinaimi2015ValuesAH}. By prompting clients to engage in such questioning and fact finding, these documents could foster reflective practices among customers and lead to greater ethical competency. 

This distribution of responsibility also poses risks. As Ahmed and Swan note, it’s important to “[track] how diversity gets used within organisations by being attached to specific bodies, units or agencies” \cite{ahmed2006doing}. Here, the documents consistently push diversity work downstream, either by assigning it to customers or by instructing customers on how to assign it to their users or customers. The result is that responsibility never “attaches” to the firms, which are continually reiterated as merely providing tools, documentation, and education. Rather than an elision of responsibility, however, these firms frame these responsibility-eliding activities as being responsible. This reconfiguration of responsibility makes the landscape of possible wrongdoing more fragmented, distributed and opaque. In Ahmed’s terms, responsibility is “refused and diffused” \cite{ahmed2007language}. %(p. 251). 
This diffusion of responsibility--beyond the responsibility for passing responsibility on--is in tension with human rights norms which direct businesses to take responsibility for avoiding and mitigating human rights impacts “directly linked to their operations, products or services by their business relationships, even if they have not contributed to those impacts" \cite{ruggie2020social}.

\section{Conclusion}
Studying the discursive practices of leading technology companies as they seek to define and structure AI ethics work provides a window into how the practices and configurations of production are reshaping concepts, defining expertise, and allocating responsibility for the enactment and protection of diversity and inclusion on the ground. As Gurses et al. writes, “[I]nquiries into [this] production can help us better engage with new configurations of power that have implications for fundamental rights and freedoms” \cite{gurses_vanhoboken_2018}. These documents provide insight into "every day ethics", revealing how diversity and inclusion--which was transformed by managers and corporate practices to align with business goals--is being broadened yet again, and instrumentalized to accommodate product teams and engineering logics and technical workflows. 

On one hand, this reconfiguration of diversity and inclusion work is pragmatic, making it easier for engineers to put ethics into practice. On the other hand, these broadened definitions construct a world in which corporations can make progress on AI ethics without addressing diversity and inclusion issues in the workforce and broader society. As technology companies continue to receive criticism on the managerial side for their slow progress in diversity statistics across all employees since 2014 \cite{rooney2020} as well as highly reported individual incidents such as Google's 2020-2021 firing of leading AI ethics researchers Dr. Timnit Gebru and Dr. Margaret Mitchell \cite{hao2020}, they have produced and promoted an increasing number of tools and publications aimed at engineering diverse, fair, and inclusive products and services. In this reconfiguration, in the absence of a diverse workforce firms can still push forward diversity and inclusion goals in the form of diverse data sets or personalization tools that can accommodate diverse behaviors and contexts. While these latter goals and work flows may produce better customer experiences, changing data practices may very well be easier than altering hiring practices, corporate culture, and business models, which are all easier than addressing the structural inequalities that civil rights laws seek to address. Methods and tools that bring diversity and inclusion into engineering practice are valuable and necessary for many purposes, but ensuring this drift into engineering logic advances civil rights ideals requires sustained efforts and clear structures that put them in service of them.  
\begin{acks}
Thank you to the anonymous reviewers for their useful comments. A special thank you to Dr. Anna Lauren Hoffman for her participation in the research and early versions of the paper. The authors also thank the members of the UC Berkeley Algorithmic Fairness and Opacity Working Group for their thoughtful review and suggestions on a early draft of this paper.

Research for this article has been supported by generous funding from the National Science Foundation (CCESTEM: Emerging Cultures of Data Science Ethics in the Academy and Industry).

\end{acks}

\bibliographystyle{ACM-Reference-Format}
\bibliography{sample-base}
\clearpage
\onecolumn
\section*{Supplementary Materials}

\appendix
\section{List of Documents}
\begin{longtable}{|l|p{6cm}|p{7cm}|l|p{1.5cm}|}
\hline
ID & Document Name & URL & Company & Department \\ \hline
G1 & Responsible AI Practices & \url{https://ai.google/responsibilities/responsible-ai-practices} & Google & AI \\ \hline
G2 & Inclusive ML & \url{https://cloud.google.com/inclusive-ml} & Google & Cloud \\ \hline
G3 & Attacking discrimination with smarter machine learning & \url{https://research.google.com/bigpicture/attacking-discrimination-in-ml} & Google & Research \\ \hline
G4 & Model Cards & \url{https://modelcards.withgoogle.com/about} & Google & Research \\ \hline
G5 & What’s new when working with AI: Explainability and Trust & \url{https://pair.withgoogle.com/chapter/explainability-trust} & Google & PAIR \\ \hline
G6 & Playing with AI Fairness & \url{https://pair-code.github.io/what-if-tool/ai-fairness.html} & Google & PAIR \\ \hline
G7 & AI at Google: our principles & \url{https://blog.google/technology/ai/ai-principles} & Google & Principles \\ \hline
G8 & tensorflow / fairness indicators & \url{github.com/tensorflow/fairness-indicators/blob/master/fairness\_indicators/documentation/guidance.md} & Google & Tensorflow \\ \hline
G9 & A guide to machine learning (ML) fairness & \url{https://www.thinkwithgoogle.com/feature/ml-fairness-for-marketers/\#next-steps} & Google & Other \\ \hline
G10 & Building ML models for everyone: understanding fairness in machine learning & \url{https://cloud.google.com/blog/products/ai-machine-learning/building-ml-models-for-everyone-understanding-fairness-in-machine-learning} & Google & Cloud \\ \hline
G11 & Explainable AI Beta & https://cloud.google.com/explainable-ai & Google & Cloud \\ \hline
G12 & What’s new when working with AI: Errors and Graceful failure & \url{https://pair.withgoogle.com/chapter/errors-failing/} & Google & PAIR \\ \hline
G13 & People + AI Guidebook: Data Collection \& Evaluation & \url{https://pair.withgoogle.com/chapter/data-collection/} & Google & PAIR \\ \hline
G14 & Responsible AI: Putting our principles into action & \url{https://www.blog.google/technology/ai/responsible-ai-principles/} & Google & Other \\ \hline
G15 & Responsible Development of AI & \url{https://ai.google/static/documents/responsible-development-of-ai.pdf} & Google & AI \\ \hline
M1 & AI for Accessibility grants & https://www.microsoft.com/en-us/ai/ai-for-accessibility-grants?activetab=pivot1:primaryr2 & Microsoft & AI \\ \hline
M2 & AI for Accessibility & \url{https://www.microsoft.com/en-us/ai/ai-for-accessibility}& Microsoft & AI \\ \hline
M3 & Featured AI for Accessibility projects & \url{https://www.microsoft.com/en-us/ai/ai-for-accessibility-projects?activetab=pivot1:primaryr2} & Microsoft & AI \\ \hline
M4 & Mitigate unfairness in machine learning models & \url{https://docs.microsoft.com/en-us/azure/machine-learning/concept-fairness-ml} & Microsoft & Azure \\ \hline
M5 & Digital Inclusion & \url{https://blogs.microsoft.com/eupolicy/digital-inclusion/} & Microsoft & EU Policy \\ \hline
M6 & AI principles: Summary and Resources & \url{https://docs.microsoft.com/en-us/learn/modules/responsible-ai-principles/9-summary-resources} & Microsoft & Learn \\ \hline
M7 & Identify guiding principles for responsible AI & \url{https://docs.microsoft.com/en-us/learn/modules/responsible-ai-principles/4-guiding-principles} & Microsoft & Learn \\ \hline
M8 & Implications of responsible AI - Practical guide & \url{https://docs.microsoft.com/en-us/learn/modules/responsible-ai-principles/3-implications-responsible-ai-practical} & Microsoft & Learn \\ \hline
M9 & Debugging data: Microsoft researchers look at ways to train AI systems to reflect the real world & \url{https://blogs.microsoft.com/ai/debugging-data-microsoft-researchers-look-ways-train-ai-systems-reflect-real-world} & Microsoft & Blog \\ \hline
M10 & Guidelines for responsible implementation of Personalizer & \url{https://docs.microsoft.com/en-us/azure/cognitive-services/personalizer/ethics-responsible-use} & Microsoft & Azure \\ \hline
M11 & Six Principles for Developing and Deploying Facial Recognition Technology & \url{https://blogs.microsoft.com/wp-content/uploads/prod/sites/5/2018/12/MSFT-Principles-on-Facial-Recognition.pdf} & Microsoft & Blog \\ \hline
M12 & Fairness in Machine learning & \url{fairlearn.github.io/user\_guide/fairness\_in\_machine\_learning.html} & Microsoft & Fairlearn \\ \hline
M13 & In Pursuit of Inclusive AI & \url{https://medium.com/microsoft-design/in-pursuit-of-inclusive-ai-eb73f62d17fc} & Microsoft & Design \\ \hline
M14 & Responsible AI: inclusiveness & \url{https://www.microsoft.com/en-us/ai/responsible-ai} & Microsoft & AI \\ \hline
M15 & Responsible bots: 10 guidelines for developers of conversational AI & \url{https://www.microsoft.com/en-us/research/publication/responsible-bots/}& Microsoft & Research \\ \hline
M16 & Age of Intelligence & \url{https://news.microsoft.com/wp-content/uploads/prod/sites/45/2019/02/Microsoft-AI-Whitepaper-Age-of-Intelligence.pdf}& Microsoft & News \\ \hline
S1 & Bias in AI: Join a Conversation with Team Einstein & \url{https://www.salesforce.com/blog/2019/04/bias-in-ai-einstein-interview.html} & Salesforce & Einstein AI \\ \hline
S2 & Ethical frameworks, tool kits, principles, and oaths - Oh & \url{https://blog.einstein.ai/frameworks-tool-kits-principles-and-oaths-oh-my/} & Salesforce & Einstein AI \\ \hline
S3 & ERASER: A Benchmark to Evaluate Rationalized NLP Models & \url{https://blog.einstein.ai/eraser-a-benchmark-to-evaluate-rationalized-nlp-models/} & Salesforce & Einstein AI \\ \hline
S4 & AI ethics & \url{https://einstein.ai/ethics} & Salesforce & Einstein AI \\ \hline
S5 & Salesforce Experience and Design & \url{https://medium.com/salesforce-ux/how-to-build-ethics-into-ai-part-i-bf35494cce9} & Salesforce & Salesforce UX \\ \hline
S6 & Understand the Ethical Use of Technology & \url{https://trailhead.salesforce.com/en/content/learn/modules/responsible-creation-of-artificial-intelligence/understand-the-ethical-use-of-technology} & Salesforce & Trailhead \\ \hline
S7 & Learn the Basics of Artificial Intelligence & \url{https://trailhead.salesforce.com/en/content/learn/modules/responsible-creation-of-artificial-intelligence/learn-the-basics-of-ai} & Salesforce & Trailhead \\ \hline
S8 & Recognize Bias in Artificial Intelligence & \url{https://trailhead.salesforce.com/en/content/learn/modules/responsible-creation-of-artificial-intelligence/recognize-bias-in-ai} & Salesforce & Trailhead \\ \hline
% S9 & Acceptable Use and External-Facing Services Policy & \url{https://www.salesforce.com/content/dam/web/en\_us/www/documents/legal/Agreements/policies/ExternalFacing\_Services\_Policy.pdf} & Salesforce & Fill in \\ \hline
S9 & How to Build Ethics into AI - Part II & \url{blog.einstein.ai/how-to-build-ethics-into-ai-part-iiresearch-based-recommendations-to-keep-humanity-in-ai} & Salesforce & Salesforce UX\\ \hline
S10 & Ethical and Humane Use of Technology & \url{salesforce.org/about-us/ethical-humane-use-of-technology} & Salesforce & Einstein AI \\ \hline
S11 & Remove Bias from Your Data and Algorithms & \url{https://trailhead.salesforce.com/en/content/learn/modules/responsible-creation-of-artificial-intelligence/remove-bias-from-your-data-and-algorithms}& Salesforce & Trailhead \\ \hline
S12 & 3 Ways to Counter Unconscious Bias in AI & \url{https://www.salesforce.com/blog/2018/09/ways-to-counter-unconscious-bias-ai.html} & Salesforce & Blog \\ \hline
S13 & View the Privacy and Ethical Use Principles Guiding our COVID-19 Response & \url{https://www.salesforce.com/content/dam/web/en\_us/www/documents/legal/Privacy/privacy-and-ethical-use-principles-guiding-our-covid-19-response.pdf} & Salesforce & Legal \\ \hline
S14 & 32 Smart Artificial Intelligence Quotes from the Experts & \url{https://www.salesforce.com/blog/2019/04/ai-quotes.html} & Salesforce & Blog \\ \hline
S15 & Why Ethical AI Is a Mindset, Not a Checklist & \url{https://www.salesforce.com/blog/2018/11/ethical-ai-is-a-mindset-not-a-checklist.html} & Salesforce & Blog \\ \hline
\end{longtable}
\newpage
\section{Fairness Analytic}
\begin{table}[h!]
 \noindent\begin{tabularx}{\textwidth}{||X|X|X|X||}
  \hline \hline
 \textbf{Dimension of \mbox{Fairness}} & \textbf{Description of \mbox{Dimension}} & \textbf{Example of \mbox{Dimension}} & \textbf{Interrogation \mbox{Questions}} \\ \hline \hline
 \multicolumn{4}{||c||}{\textsc{Dimensions of Protection}}\\ \hline \hline
 Target & The ideal end state toward which fairness aspires. At a high level, this could be substantive or procedural. &
 \vspace{-6pt}
 \begin{itemize}[leftmargin=*, noitemsep, topsep=0pt, label={$\circ$}]
 \item Formal equality (blind to all other variables)---to each person an equal share;
 \item Need-based allocation---to each person according to individual need;
 \item Effort-based allocation---to each person according to individaul effort;
 \item Social contribution---to each person according to societal contribution;
 \item Merit-based allocation---to each person according to merit;
 \item Information and participation rights
 \item Accurate and robust representation
 \end{itemize} & What should fairness provide?\\ \hline
 %\multirow{2}{0.22\textwidth}
 {Subject (and, in relation to who/what?)} & Actor(s) or Entity(ies) to whom fairness is provided. &
 \vspace{-6pt}
 \begin{itemize}[leftmargin=*, noitemsep, topsep=0pt, label={$\circ$}]
 \item Individual
 \item Social Groups
 \item Roles
 \end{itemize} &
 Fairness is at stake for whom or what?\\ 
 & Fairness is often used comparatively, requiring the construction of categories along some attribute or set of attributes.
 & 
 & What properties or attributes are being made fair? What groups are being compared? Granularity? \\\hline
% \end{tabularx}
% \caption{Dimensions of protection for contests over fairness.}
% \end{table}

%  \noindent\begin{tabularx}{\textwidth}{||X|X|X|X||}
%   \hline \hline

 \multicolumn{4}{||c||}{\textsc{Dimensions of Provision}}\\ \hline \hline
 Provider & Actor(s) charged with being fair or avoiding unfairness. &
 \vspace{-6pt}
 \begin{itemize}[leftmargin=*, noitemsep, topsep=0pt, label={$\circ$}]
 \item Government
 \item Business entity
 \item Technology
 \item Individuals
 \end{itemize} & Who or what is supposed to behave fairly or avoid unfair behavior?\\ \hline
 Mechanism & Modalities used to support fairness. &
 \vspace{-6pt}
 \begin{itemize}[leftmargin=*, noitemsep, topsep=0pt, label={$\circ$}]
 \item Legal regulations
 \item Technical design
 \item Business processes
 \item Education
 \item Norms
 \end{itemize} & How is fairness operationalized?\\ \hline
 Implementer & Actor(s) tasked with operationalizing fairness through chosen modalities. &
 \vspace{-6pt}
 \begin{itemize}[leftmargin=*, noitemsep, topsep=0pt, label={$\circ$}]
 \item Lawyers
 \item Engineers
 \item Product Managers
 \item Designers
 \item Professional associations
 \item Educators
 \end{itemize} & Who brings fairness into practice?\\ \hline \hline
 \end{tabularx}
 \end{table}
 \clearpage
\twocolumn

\end{document}